\documentclass{easychair}

\usepackage{doc}
\usepackage{subcaption}

\newcommand{\milkmaid}{\texttt{MeLoId}}
\newcommand{\psl}{\texttt{PSL}}

\newcommand{\pamper}{\texttt{PaMpeR}}
\newcommand{\conjecture}{\texttt{Conjecture}}
\newcommand{\fastforce}{\texttt{Fastforce}}
\newcommand{\quickcheck}{\texttt{Quickcheck}}
\newcommand{\false}{\texttt{False}}
\newcommand{\dynamic}{\texttt{Dynamic}}
\newcommand{\induct}{\texttt{induct}}
\title{Towards Machine Learning Induction %for Isabelle/HOL%
}

\author{
Yutaka Nagashima\inst{1}\inst{2}
}

\institute{
  CIIRC, Czech Technical University in Prague,\\
  Prague, Czech Republic
\and
  Department of Computer Science, University of Innsbruck,\\
  Innsbruck, Tyrol, Austria
 }

\authorrunning{Nagashima}

\titlerunning{Towards Machine Learning Induction}

\begin{document}

\maketitle

\begin{abstract}
  Induction lies at the heart of mathematics and computer science.
  However, automated theorem proving of inductive problems is still limited in its power.
  In this abstract, we first summarize our progress in automating inductive theorem proving
  for Isabelle/HOL.
%  Then, we identify three major obstacles 
%  that arise when applying machine machine learning to automated inductive theorem proving.
%  Finally, 
  Then, we present \milkmaid{}, 
  our approach to suggesting 
  promising applications of induction without completing a proof search.
\end{abstract}

% The table of contents below is added for your convenience. Please do not use
% the table of contents if you are preparing your paper for publication in the
% EPiC Series or Kalpa Publications series

%\setcounter{tocdepth}{2}
%{\small
%\tableofcontents}

%\section{To mention}
%
%Processing in EasyChair - number of pages.
%
%Examples of how EasyChair processes papers. Caveats (replacement of EC
%class, errors).

%------------------------------------------------------------------------------
\section{PSL and Goal-Oriented Conjecturing for Isabelle/HOL}
\label{sect:PSL_PGT}
Previously, we developed \psl{} \cite{psl} %, a Proof Strategy Language, 
for Isabelle/HOL \cite{tutorial}
and its extension to conjecturing mechanism \cite{pgt}
as initial steps towards the development of a smart proof search in Isabelle \cite{nagashima2017smart}.
With \psl{} one can write the following strategy for induction:

\begin{verbatim}
strategy DInd = Thens [Dynamic (Induct), Auto, IsSolved]
\end{verbatim}

\noindent
\psl{}'s \verb|Dynamic| keyword creates variations of the \verb|induct| method
by specifying different combinations of promising arguments
found in the proof goal and its background proof context.
Then, \verb|DInd|
combines these induction methods with the general purpose proof method, \verb|auto|,
and \verb|is_solved|,
which checks if there is any proof goal left after applying \verb|auto|.
%As shown in Fig. \ref{fig:psl},
\psl{} keeps applying the combination of
a specialization of \verb|induct| method and \verb|auto|,
until either \verb|auto| discharges all remaining sub-goals or
\verb|DInd| runs out of the variations of \verb|induct| methods.
%as shown in Fig.~\ref{fig:psl}.

Sometimes it is necessary for human-engineers to come up with auxiliary lemmas,
from which they can derive the original goal.
To automate this process, we developed a new atomic strategy, \conjecture{}, 
as an extension to \psl{}.
Given a proof goal, \conjecture{} first produces various conjectures that might be useful
to discharge the original proof goal, 
then inserts these conjectures as the premise of the original goal.
Thus, for each conjecture, \psl{} produces two sub-goals: 
the first sub-goal states that the conjecture implies the original goal, 
and the second sub-goal states that the conjecture indeed holds.
With \conjecture{} integrated into \psl{}, one can write the following strategy:

\begin{verbatim}
strategy CDInd = Thens [Conjecture, Fastforce, Quickcheck, DInd]
\end{verbatim}

\noindent
The sequential application of \fastforce{} prunes
conjectures that are not strong enough to prove the original goal,
whereas the application of \quickcheck{} attempts to prune 
conjectures that are equivalent to \false{}. %, 
%narrowing the search space for \verb|DInd|.% as shown in Fig.~\ref{fig:pgt}.
% \begin{figure}[h]
%   \centering
%   \begin{subfigure}[]{0.4\textwidth}
%       \includegraphics[width=\textwidth]{image/psl_search_tree.pdf}
%       \caption{Search tree of \texttt{DInd}}
%       \label{fig:psl}
%   \end{subfigure}
%   ~ %add desired spacing between images, e. g. ~, \quad, \qquad, \hfill etc.
%     %(or a blank line to force the sub-figure onto a new line)
%   \begin{subfigure}[]{0.4\textwidth}
%       \includegraphics[width=\textwidth]{image/pgt_search_tree.pdf}
%       \caption{Search tree of \texttt{CDInd}}
%       \label{fig:pgt}
%   \end{subfigure}
%   \caption{\psl{}'s proof search with/without \pgt{}.}
%   \label{fig:example_search_tree}
% \end{figure}
This way, we can narrow the search space by focusing on promising conjectures;
however, when proof goals require many applications of inductions and multiple conjecturing steps,
the search space blows up rapidly due to the various \induct{} methods produced by
the \dynamic{} keyword.
Since the \induct{} method usually preserves the provability of proof goal,
even when the \induct{} method has arguments that are inappropriate to discharge the proof goal,
counter-example finders, such as \quickcheck{}, cannot discard them.
To address this problem,
we are developing \milkmaid{} to suggest how to apply induction
without completing a proof.
%The missing piece necessary to automate inductive theorem proving is the tool to decide
%how to apply mathematical induction without discharging the remaining sub-goals.

%------------------------------------------------------------------------------
\section{\milkmaid{}: Machine Learning Induction}
\label{sect:MiLkMaId}

%\paragraph{Overview.}
%Fig. \ref{fig:milkmaid} 
The figure below illustrates the overall architecture of \milkmaid{}.
Similarly to \pamper{} \cite{pamper},
which suggests promising proof methods 
for a given proof goal and its underlying context,
\milkmaid{} tries to learn how to apply induction effectively using 
human-written proof corpora as training data. %in the preparation phase.
In the preparation phase, \milkmaid{} collects invocations of the \induct{} method
appearing in the proof corpora and converts each of them into a simpler format,
a vector of booleans using an assertion-based feature extractor.
Then, \milkmaid{} constructs a regression tree \cite{BreimanFOS84}, which describes not only
which variations of the \induct{} method are promising but also
which assertions are useful to make such recommendations in the recommendation phase.

%\paragraph{Active Mining.}
The mechanism of \milkmaid{} differs from that of \pamper{} in multiple ways.
First of all, \milkmaid{} analyzes proof corpora via what we call \textit{active mining}:
\milkmaid{} first creates various \induct{} methods with distinct combinations of arguments, 
applies each of them to the goal, and compares their results.
Secondly, the input to \milkmaid{}'s assertions are the triples of 
a goal with its context, 
the arguments to the \induct{} method, and 
the sub-goal appearing after applying \induct{},
whereas \pamper{}'s assertions consider only the first two as input.
\milkmaid{} takes the emerging sub-goals into considerations:
Since the application of the \induct{} method alone is not time-consuming,
we expect that it is desirable to improve the accuracy of recommendation using the emerging sub-goals
even at the cost of the extra time spent by the \induct{} method.
Third, \milkmaid{} assertions tend to analyze the structures of the triples,
while \pamper{}'s assertions tend to focus on the names of constants and types appearing in the 
proof goal at hand.

We have implemented the active mining mechanism and around 40 assertions.
Our preliminary experiment suggests that
the feature extractor successfully distills the essence of
some undesirable combinations of arguments of \induct{}.
%the feature extractor started to differentiate 
%some inappropriate applications of the \induct{} method;
However, more comprehensive evaluation and further engineering efforts remain as our future work.

\begin{figure}[b]
\centering
%	\begin{centering}
	\includegraphics[width=0.9\textwidth]{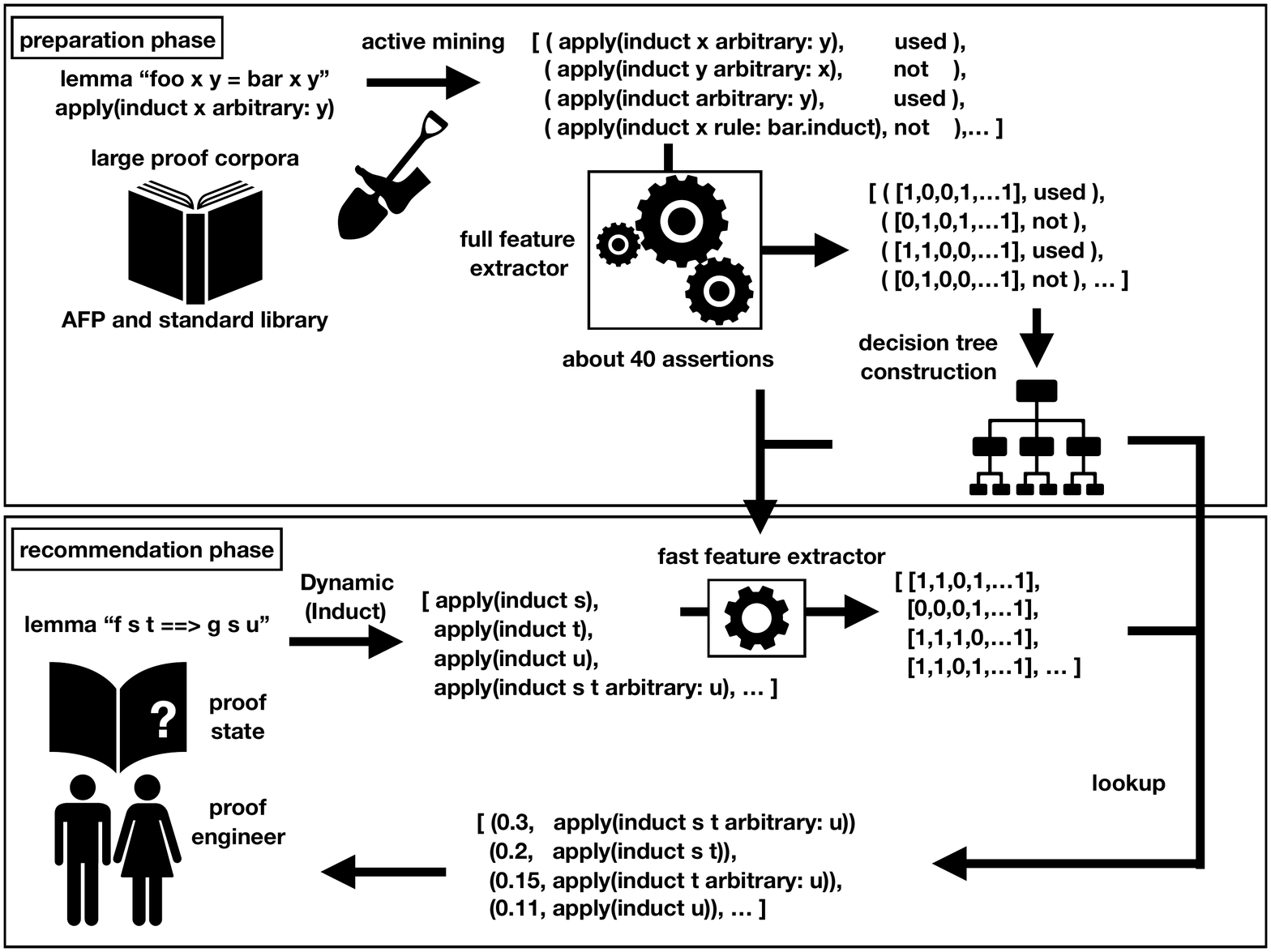}
%	\caption{\milkmaid}
%	\end{centering}
	\label{fig:milkmaid}
\end{figure}
%------------------------------------------------------------------------------
\newpage
\section{Acknowledgments}
\label{sect:acks}
This work was supported by the European Regional Development Fund under the project 
AI \& Reasoning (reg. no.CZ.02.1.01/0.0/0.0/15\_003/0000466).

%------------------------------------------------------------------------------
\label{sect:bib}
\bibliographystyle{plain}
\bibliography{easychair}
\newpage

%------------------------------------------------------------------------------
% Index
%\printindex

%------------------------------------------------------------------------------
\end{document}